\documentclass{emulateapj}

\begin{document}

\shortauthors{Luhman}
\shorttitle{Wide Binary in Upper Scorpius}

\title{Discovery of a Wide Low-mass Binary System in Upper Scorpius}

\author{K. L. Luhman\altaffilmark{1}}
\affil{Department of Astronomy and Astrophysics,
The Pennsylvania State University, University Park, PA 16802;
kluhman@astro.psu.edu.} 

\altaffiltext{1}
{Visiting Astronomer at the Infrared Telescope Facility, which is operated
by the University of Hawaii under Cooperative Agreement no. NCC 5-538 with
the National Aeronautics and Space Administration, Office of Space Science,
Planetary Astronomy Program.}

\begin{abstract}

Using the near-infrared spectrometer SpeX and its slit-viewing camera at 
the IRTF, I have resolved a low-mass member of the Upper Scorpius OB 
association into a double star.
From $K$-band images of the pair, DENIS-P~J161833.2-251750.4~A and B, 
I measure a separation of $0\farcs96$ and a magnitude difference of
$\Delta K=0.42$~mag. 
I present resolved 0.8-2.5~\micron\ spectroscopy of the two objects,
both of which exhibit signatures of youth in the shape of their 
$H$- and $K$-band continua, demonstrating that both are members
of Upper Scorpius rather than field stars. In addition, through a comparison
to optically-classified pre-main-sequence objects, I derive a spectral type
near M5 for each component, corresponding to a mass of $\sim0.15$~$M_{\odot}$ 
with the evolutionary models of Chabrier and Baraffe.
The probability that this pair is composed of unrelated M-type 
members of Upper Scorpius is $\sim10^{-5}$.
When added to the recent discoveries of other wide, easily disrupted
low-mass binaries, this new system further establishes that the formation
of low-mass stars and brown dwarfs does not require ejection from multiple
systems. These observations also indicate that
wide low-mass binaries can form in OB associations as well as in smaller 
clusters where the previously known wide pairs have been found.
Thus, the available data show no perceptible effect of 
star-forming environment on the prevalence of loosely bound low-mass systems.

\end{abstract}

\keywords{infrared: stars --- stars: evolution --- stars: formation --- stars:
low-mass, brown dwarfs --- binaries: visual -- stars: pre-main sequence}

\section{Introduction}
\label{sec:intro}

The multiplicity of stars is influenced by the specific characteristics
of the star formation process, and thus has been the subject of 
extensive measurements.
Similarly, in an attempt to gain insight into the formation of brown dwarfs,
recent studies have heavily scrutinized the binary properties of objects 
near and below the hydrogen burning mass limit in the solar neighborhood
\citep{koe99,mar99,giz03,rei01,clo02a,clo02b,clo03,bou03,bur03,fre03,sie05,for05,liu05,gel05},
open clusters \citep{mar98,mar00,mar03}, and young clusters and associations 
\citep{neu02,bou04,cha04,cha05,luh04bin,luh05,kra05}.
Most of the binary low-mass stars and brown dwarfs in these surveys
exhibit maximum separations of $a\sim20$~AU, 
which is consistent with the predictions of the embryo ejection scenario
for the formation of low-mass bodies
\citep{rc01,bos01,bat02}. However, a small but growing number of 
low-mass binaries have been found at wider projected separations that
range from 33-41~AU \citep{har74,mar00,cha04,pha05,mam05} to beyond 
200~AU \citep{giz00,luh04bin,bil05}.
Because these wide binaries are weakly bound and extremely fragile, it 
is unlikely they that have been subjected to violent dynamical interactions, 
indicating that some low-mass stars and brown dwarfs are able
to form without the involvement of ejection. In fact, the role of ejection
may be minor based on other aspects of the multiplicity of low-mass objects 
\citep{mj05}.

To fully utilize multiplicity as a probe of the formation of low-mass 
bodies, it is necessary to consider populations spanning a range of ages
and star-forming conditions. OB associations represent an 
area of this phase space that remained unexplored until \citet{kra05}
obtained $HST$ images of low-mass members of Upper Scorpius,
which has an age of 5~Myr \citep{deg89,pz99} and is the nearest OB 
association at a distance 145~pc \citep{dez99}.
Among their 12 targets, they discovered three binaries, all with 
projected separations less than 18~AU.
\citet{kra05} tentatively suggested that this absence of wide systems 
in Upper Sco relative to lower density clusters like Chamaeleon~I and
TW Hya \citep{cha04,luh04bin} might indicate a variation in the formation 
process of low-mass binaries with star-forming environment.

In this Letter, I report the discovery of a widely-separated low-mass 
binary star in Upper Sco, which was found serendipitously 
during a spectroscopic survey of the low-mass members of this association.
I present near-infrared (IR) images and spectroscopy of the components of
the pair, measure their spectral types and assess their membership 
in a binary system, and discuss the implications of this new pair.

\section{Observations}
\label{sec:obs}

While conducting a spectroscopic survey of known low-mass members of the 
Upper Sco association on 2005 June 16 with SpeX \citep{ray03} at the
NASA Infrared Telescope Facility (IRTF), I found that one of the targets,
DENIS-P~J161833.2-251750.4 \citep{mar04}, 
appeared as a $\sim1\arcsec$ double on the slit-viewing camera. 
To further investigate the nature of this pair, I performed the 
imaging and spectroscopy described in this section.

\subsection{Photometry}

\begin{figure}
\plotone{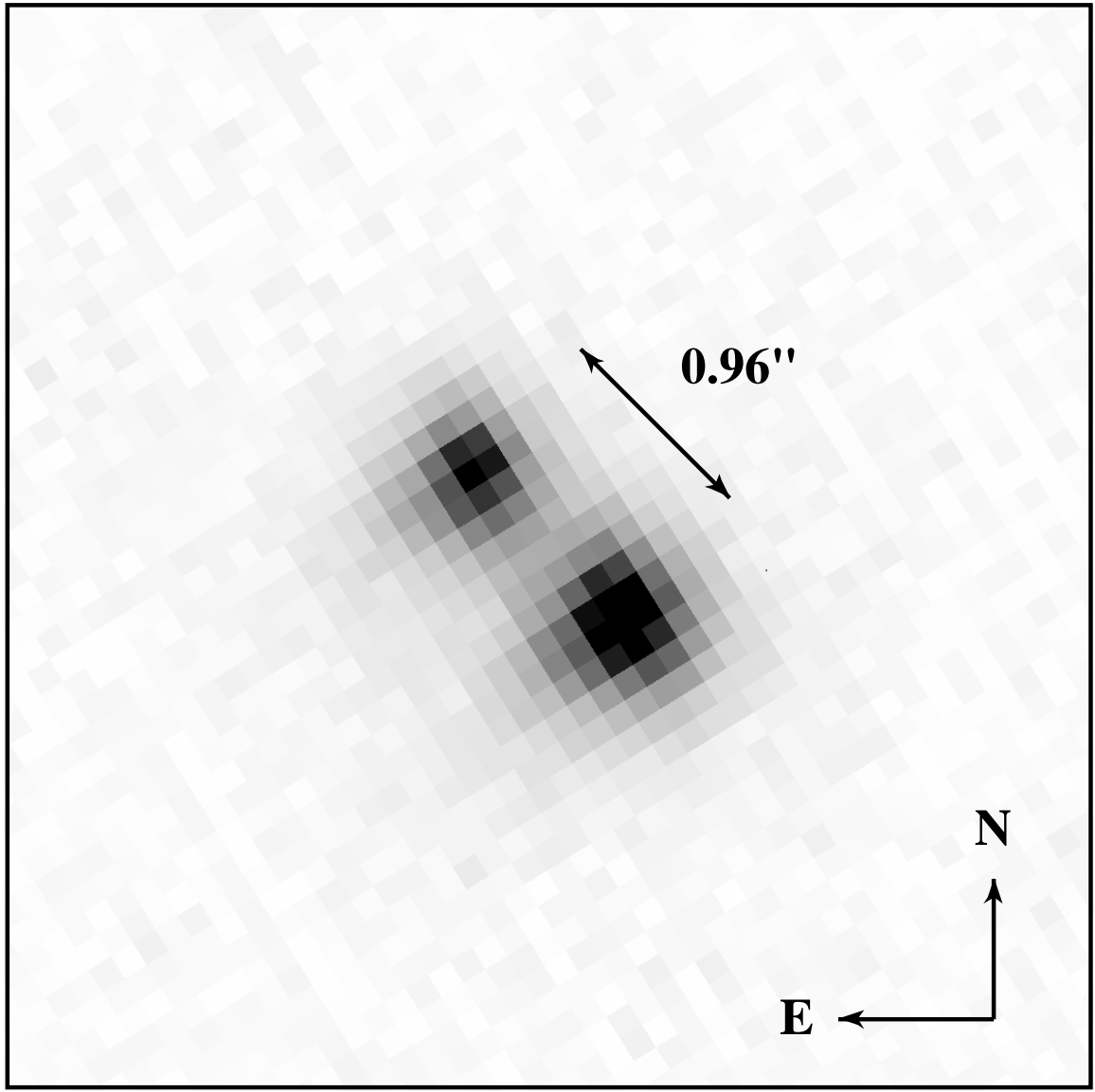}
\caption{ 
$K$-band $5\arcsec\times5\arcsec$ image of the binary system 
DENIS-P~J161833.2-251750.4 obtained with the slit-viewing camera
on SpeX (FWHM$=0\farcs5$).
}
\label{fig:image}
\end{figure}

To measure the flux ratio of the components of DENIS-P~J161833.2-251750.4
(hereafter DENIS~1618-2517),
I used the slit-viewing camera on SpeX. This camera contained a
$512\times512$ Aladdin 2 InSb array and had a plate scale of
$0\farcs12$~pixel$^{-1}$ \citep{ray03}.
Through a $K$ filter, I obtained 18 dithered one-second exposures. 
These images were median combined to produce a flat field image, which
was then divided into each original exposure. The 12 frames with the 
best image quality were registered and combined. A $5\arcsec\times5\arcsec$
subsection of the resulting image surrounding DENIS~1618-2517
is shown in Figure~\ref{fig:image}. The components of this pair exhibit 
FWHM$=0\farcs5$ in this image.  I extracted photometry for each object with 
the task PHOT under the IRAF package APPHOT using a radius of three pixels,
arriving at a magnitude difference of $\Delta K=0.42\pm0.03$~mag. 
To measure astrometry for the pair, I first determined the rotation
between the cardinal directions and the array axes using the 2MASS 
coordinates of DENIS~1618-2517 and the one other 2MASS source
appearing in the image (2MASS 16183428-2518067).
With this information, I measured a 
position angle of $50.3\pm1\arcdeg$ for the secondary relative to the
primary. Using the plate scale from \citet{ray03}, I then measured
a separation of $0\farcs96\pm0\farcs04$ for the pair.

\subsection{Spectroscopy}

Spectra were obtained of the components of DENIS~1618-2517
using SpeX in the prism mode with a $0\farcs5$ slit. This configuration 
produced a wavelength coverage of 0.8-2.5~\micron\ and a resolution of
$R\sim200$. After adjusting the position angle of the slit to align it
with the axis connecting the pair, I selected an integration time of 30~sec
and obtained a total of 18 exposures during a sequence of dithers between
two positions on the slit. The 10 frames with the best image quality were
reduced with the Spextool package \citep{cus04}.
The data were corrected for telluric absorption with the method described 
by \citet{vac03}.

\section{Analysis}
\label{sec:analysis}

\subsection{Spectral Classification}

\begin{figure}
\epsscale{.40}
\plotone{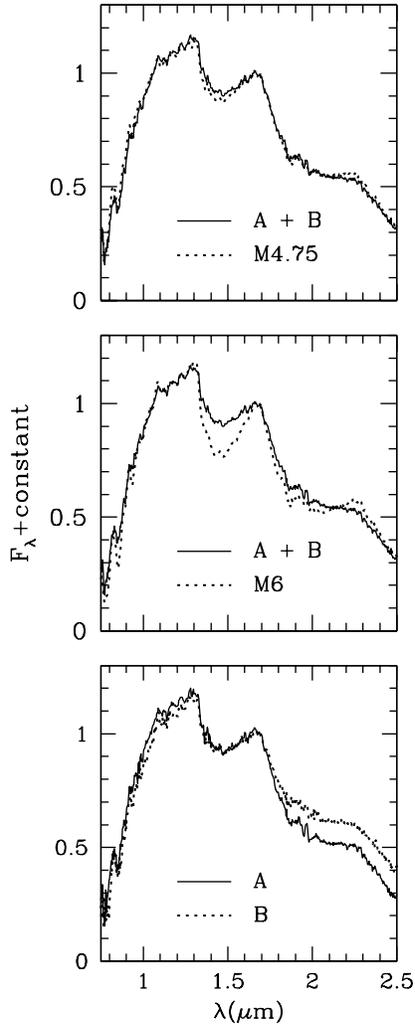}
\caption{ 
Near-IR spectrum of both components of DENIS~1618-2517
compared to spectra of Taurus members MHO~6 ({\it top}) and MHO~5 
({\it middle}). The latter sources have optical spectral types of 
M4.75 and M6 \citep{bri02} and have been reddened by $A_V=2.2$ and 1.7
to match the overall slope of DENIS~1618-2517~A+B
The resolved spectra of A and B exhibit similar spectral types ({\it bottom}).
The spectra are normalized at 1.68~\micron.
}
\label{fig:spec}
\end{figure}

I now use the photometry and spectroscopy from the previous section 
to examine the spectral types and extinctions of the components of 
DENIS~1618-2517. \citet{mar04} found that the H$\alpha$ emission and Na~I 
absorption in an unresolved optical spectrum of the pair were indicative of 
youth and thus membership in Upper Sco. 
In the SpeX data, I find evidence of youth in both objects. 
As shown in Figure~\ref{fig:spec}, each spectrum exhibits a triangular 
$H$-band continuum, which is a distinguishing characteristic of cool
pre-main-sequence objects \citep{luc01}.
The depths of the optical and IR molecular absorption bands, primarily TiO, VO, 
and H$_2$O, are also very similar between the two objects, demonstrating that
the two objects have the same spectral type to within $\sim0.5$~subclass.
In comparison, two young low-mass stars at the same age
with a $K$-band magnitude difference of 0.42 are predicted to have 
$\Delta T\sim70$~K \citep{bar98}, which corresponds to $\sim0.5$~subclass 
\citep{luh03}.
Thus, the relative spectral types and magnitudes of the components of
DENIS~1618-2517 are consistent with the coevality expected for a binary system.

Although the absorption band strengths agree between the two objects, 
the secondary is redder beyond 1.7~\micron. This is
difficult to explain as a difference in spectral type or extinction, and
instead probably is due to excess emission from circumstellar material
around the secondary, an imperfect separation of the light from the two 
objects during extraction of their spectra, or the combination of 
misalignment of the slit along the binary axis and differential refraction.
Meanwhile, the IR spectra of both DENIS~1618-2517~A and B are significantly 
redder than most of the other low-mass members of Upper Sco that were observed 
during the same run. The same trend is present in the 2MASS $J-H$ and 
$H-K_s$ colors of these objects and is consistent with extinction. 
Relative to the bluest (i.e., unreddened) SpeX data of members of Upper Sco 
and other young clusters at M5-M6 \citep{luh05}, the spectrum
of DENIS~1618-2517~A+B exhibits an extinction of $A_V\sim3$. 
The presence of significant reddening is additional evidence that these
objects are not foreground field 
dwarfs.\footnote{The relatively high extinction toward DENIS~1618-2517
could be caused by local circumstellar dust or cirrus clouds in the general
vicinity of DENIS~1618-2517 \citep{mey93}.}
Based on the strength of the molecular absorption bands,
DENIS~1618-2517~A and B are earlier than most of the other 
Upper Sco members observed in this run, which have optical spectral types 
that range from M5 to M9 \citep{ard00,mar04}. This suggests that 
DENIS~1618-2517~A+B may be closer to M5 than the value of M6 measured
by \citet{mar04}. I find the same result when I compare 
the pair to optically-classified members of other young clusters. 
For instance, as shown in Figure~\ref{fig:spec}, the molecular absorption
bands of DENIS~1618-2517~A+B agree better with those of MHO~6 than with
MHO~5, which have optical types of M4.75 and M6, respectively \citep{bri02}.
The relatively large reddening toward DENIS~1618-2517 may have caused
\citet{mar04} to arrive at a spectral type that was slightly too late. 
Combining a spectral type of M5 with the temperature scale of \citet{luh03}
and the evolutionary models of \citet{bar98} implies a mass of
$\sim0.15$~$M_\odot$ for each component of DENIS~1618-2517.

\subsection{Evidence of Binarity}

The results from the previous section and from \citet{mar04} 
demonstrated that the components of DENIS~1618-2517 are young members
of Upper Sco. I now examine if they are likely to comprise a true binary system
rather than a pair of unrelated members of the association that happen 
to have a small projected separation.
The surface density of M-type members of Upper Sco is $\sim5$~deg$^{-2}$
and the total area of the association is $\sim100$~deg$^2$
\citep{ard00,pre01,pre02,mar04}.
Across the entire association, the probability of finding a pair of 
M-type members with a projected separation of $a\leq1\arcsec$ is 
$3\times10^{-4}$.
The observations that discovered this pair considered only a small fraction of 
the membership (a few dozen) and were not designed to detect close
pairs. As a result, the relevant probability for DENIS~1618-2517 is lower
by at least an order of magnitude. Based on the very low value of this 
probability, I conclude that DENIS~1618-2517~A and B comprise a binary system 
rather than two unrelated members of Upper Sco.
Using the average distance of 145~pc for the association \citep{dez99}, 
the projected separation of $0\farcs96$ of this pair corresponds to 140~AU.

Demonstrating a common proper motion for a pair of stars
is the conventional method of conclusively establishing binarity in the
solar neighborhood, but it is less useful for young populations because
even unrelated cluster members are comoving at the level of precision
typically available for these measurements.  For instance, 
the common proper motions measurements for the recently discovered substellar
companions to GQ~Lup \citep{neu05} and 2M~1207-3932 \citep{cha05}
are not precise enough to distinguish between true binaries and 
comoving unbound cluster members seen in projection near each other. 
Thus, for GQ~Lup, 2M~1207-3932, previously known
pairs in Upper Sco \citep{kra05}, and indeed the vast majority of visual 
doubles in young clusters and associations, available evidence of binarity
is based on the same statistical 
considerations presented in this work for DENIS~1618-2517. 
Only the value of the probability differs among these
binaries, which depends on the separation of a given pair and the surface 
density of cluster members. The probability of unrelated 
clusters members with small separations is exceedingly low for sparse
populations like Upper Sco, Chamaeleon, Lupus, and TW Hya 
\citep{kra05,luh04bin,neu05,cha04} but non-negligible for dense 
clusters like the Trapezium in Orion \citep{pro94}.

\section{Discussion}
\label{sec:disc}

As with the brown dwarf pair in Chamaeleon from \citet{luh04bin}, 
the binarity of DENIS~1618-2517 was discovered serendipitously
and thus cannot be used at this time in a multiplicity measurement for 
low-mass members of Upper Sco. However, the mere existence of this binary 
system has important implications. 
First, DENIS~1618-2517 is one of only a few known low-mass binaries
with very wide separations \citep[$a>100$~AU,][]{giz00,luh04bin,bil05}. 
These wide pairs are fragile and easily disrupted, and thus demonstrate
that some low-mass stars and brown dwarfs can form without the assistance
of ejection.
In addition, with these observations of DENIS~1618-2517, wide low-mass
binaries are now known across a wider range of star-forming conditions,
from small, sparse clusters (Chamaeleon) and to rich OB associations 
(Upper Sco).
Thus, in the measurements available to date, the formation of wide low-mass
binaries has no discernible dependence on star-forming environment.

\acknowledgements
K. L. was supported by grant NAG5-11627 from the NASA Long-Term Space
Astrophysics program.

\end{document}